\documentclass[%
reprint,
showpacs,
amsmath,amssymb,
aps,
prl,
]{revtex4-1}

\usepackage{graphicx}% Include figure files
\usepackage{dcolumn}% Align table columns on decimal point
\usepackage{bm}% bold math
\usepackage{braket}
\usepackage{color}
\usepackage{cancel}
\usepackage[normalem]{ulem}

\begin{document}

\preprint{APS/123-QED}

\title{How to distinguish squeezed and coherent phonons in femtosecond x-ray diffuse scattering}
%\title{How to distinguish coherent and squeezed phonons}
\author{T. Henighan$^{1,2}$}\thanks{henighan@slac.stanford.edu}
\author{M. Trigo$^{1,3}$}
\author{M. Chollet$^4$}
\author{J. N. Clark$^{1,5}$}
\author{S. Fahy$^{6,7}$}
\author{J. M. Glownia$^4$}
\author{M. P. Jiang$^1$}
\author{M. Kozina$^1$}
\author{H. Liu$^1$}
\author{S. Song$^4$}
\author{D. Zhu$^4$}
\author{D. A. Reis$^{1,3,8}$}\thanks{dreis@slac.stanford.edu}

\affiliation{$^1$PULSE Institute, SLAC National Accelerator Laboratory, Menlo Park, California, USA}

\affiliation{$^2$Physics Department, Stanford University, Stanford, California, USA}

\affiliation{$^3$SIMES Institute, SLAC National Accelerator Laboratory, Menlo Park, California, USA} 

\affiliation{$^4$Linac Coherent Light Source, SLAC National Accelerator Laboratory, Menlo Park, California, USA}

\affiliation{$^5$Center for Free-Electron Laser Science, Deutsches Elektronensynchrotron, 22607 Hamburg, Germany}

\affiliation{$^6$Tyndall National Institute, Cork, Ireland}

\affiliation{$^7$Department of Physics, University College Cork, Cork, Ireland}

\affiliation{$^8$Department of Photon Science and Applied Physics, Stanford University, Stanford, California, USA}

\date{\today}

\begin{abstract}

Impulsive optical excitation can generate both coherent and squeezed phonons.  The expectation value of the phonon displacement $\braket{u_q}$ oscillates at the mode frequency for the coherent state but remains zero for a pure squeezed state.  In contrast, both show oscillations in 
$\braket{|u_q|^2}$ at twice the mode frequency.  Therefore it can be difficult to distinguish them in a second-order measurement of the 
displacements, such as in first-order x-ray diffuse scattering.  Here we demonstrate a simple method to distinguish squeezed from coherent atomic motion by measurement of the diffuse scattering following double impulsive excitation.  We find that femtosecond optical excitation generates squeezed phonons spanning the Brillouin zone in Ge, GaAs and InSb. Our results confirm the mechanism suggested in 
[Nature Physics 9, 790 (2013)].  

\end{abstract}


\maketitle
%%%%%%%%%%%%%%%%%%
%% INTRODUCTION %%
%%%%%%%%%%%%%%%%%%

Light scattering in solids provides information about low-lying excitations, such as phonons \cite{lss1, lss4, lss9}. Optical light-scattering couples to low net 
momentum excitations because the wavelength of the light is small compared to the interatomic distances.  In the case of first-order Raman scattering in perfect crystals, conservation of energy and crystal momentum dictates 
that the inelastically scattered photons involve the absorption or emission of a single phonon with reduced wavevector $\mathbf {q}\approx$0. In second-order Raman scattering, the light couples to a broad continuum of phonon pairs with 
near equal and opposite momenta, such that the reduced wavevector of each pair is near zero, but the individual phonons have $\mathbf{q}$ spanning the Brillouin zone (BZ). In the absence of perfect crystalline order, 
optical excitation can also couple to a disorder-activated first-order continuum \cite{carles1982}.  Thus, it can  be difficult to separate first and second-order scattering by measurement of the Raman spectrum alone. 

An optical pump pulse can also excite a first-order Raman-active phonon if the pulse duration is significantly shorter than the phonon period \cite{dhar1994,merlin1997}. This generates a coherent state in which the expectation value of the phonon displacement, $\braket{u_{{\mathbf q},\lambda}}$, oscillates at the natural frequency of the mode ($\lambda$ denotes the phonon branch). As in the case of frequency domain Raman, first-order impulsive optical scattering in perfect crystals couples to modes near zone-center.  In the more general case of second-order-Raman-like processes, the short pump generates a continuum of squeezed phonon pairs for which $\braket{u_{{\mathbf q}, \lambda} u_{-{\mathbf q}, \lambda}}$=$\braket{|u_{{\mathbf q}, \lambda}|^2}$ oscillates at twice the phonon frequencies  for each ${\mathbf q}$ \cite{garrett1997}. 

Recently we have demonstrated an optical-pump/x-ray probe time-domain analog of inelastic x-ray scattering \cite{trigo2013}. In these measurements a femtosecond optical pump %\sout{generates temporal oscillations in {\color{blue}$\braket{|u_{{\mathbf q}, \lambda}|^2}$} for a continuum of modes reaching all the way to the BZ-boundary \cite{zhu2015}} 
excites a broad continuum of phonons in the sample that are probed by oscillations in the x-ray diffuse scattering as a function of momentum transfer $\mathbf{K}$ and pump-probe delay $t$. 
The scattered x-ray intensity $I({\mathbf{K}},t)$ is proportional to the energy-integrated dynamical structure factor \cite{sinha2001} and thus a weighted sum of the second-order, equal-time, correlation function $\braket{u_{{\mathbf q}, \lambda}(t) u_{-{\mathbf q}, \lambda}(t)}$ over $\lambda$ at a given $\mathbf{q}=\mathbf{K}-\mathbf{G}$ (where $\mathbf{G}$ is the closest reciprocal lattice vector) \footnote{In a thermal or number state, $\braket{|u_{\mathbf{q}, \lambda}|^2}$ is equivalent to the mean-square displacement $\braket{|u_{\mathbf{q}, \lambda}-\braket{u_{\mathbf{q}, \lambda}}|^2}$} \footnote{More generally, the energy-integrated dynamical structure factor is a weighted sum over $\braket{u_{{\mathbf q}, \lambda}(t) u_{-{\mathbf q}, \lambda'}(t)}$. The terms for which $\lambda \neq \lambda'$ which can give rise to oscillations at the sum and difference frequencies of the two modes. However, such combination modes have not been observed in the tetrahedrally bonded semiconductors studied here.}.  The diffuse scattering was seen to oscillate at twice the transverse acoustic phonon frequencies as a function of $\mathbf{q}$ throughout the BZ \cite{zhu2015} as expected for a second-order measurement in the phonon displacements. 
In \citet{trigo2013}, the oscillations were attributed to 
scattering from squeezed phonon pairs generated (by second-order scattering) from the optical pump. However,  both squeezed and coherent states would generate oscillations in the second-order correlation functions at twice the mode frequency, so that the observation of the first overtone alone in the x-ray scattering cannot distinguish the excitation mechanism. 

\begin{figure}
\centering
\includegraphics[width=3.375in]{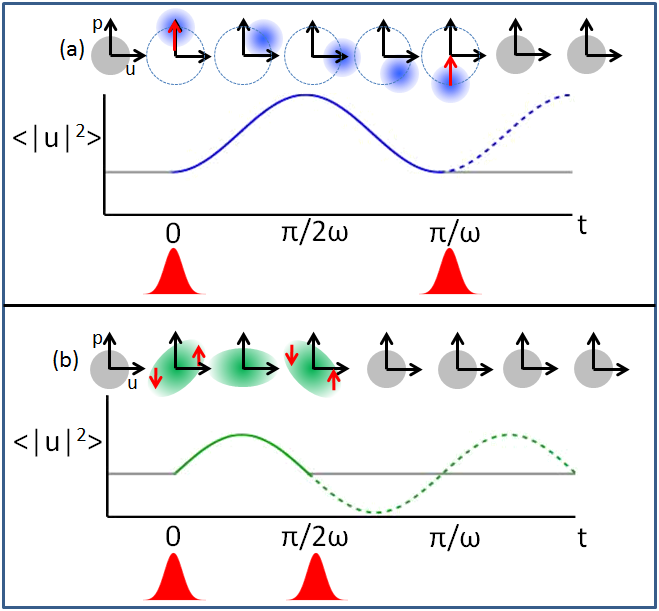}
\caption{(color online) Schematic of sudden excitation and de-excitation of coherent and squeezed modes. (a) Phase space diagram of an oscillator subjected to an impulse, generating a coherent state. An identical impulse at time $\pi/\omega$ later returns the oscillator to its original state. (b) Oscillator of the same frequency subjected to displacement-dependent impulse, generating a squeezed state. A second such displacement-dependent impulse at time $\pi/2\omega$ will return the oscillator to its original state. Note that in both cases, $\braket{|u|^2}$ oscillates at twice the natural oscillator frequency. However, the delay between the two impulsive perturbations that will return the oscillator to its original state are different for the coherent and squeezed cases. The dashed and solid lines show the evolution of $\braket{|u|^2}$ in the absence and presence of the second pump pulse, respectively.}
\label{fig:f1}
\end{figure}

Here we propose and demonstrate a method for distinguishing coherent and squeezed states in a second-order measurement in the phonon displacements using femtosecond x-ray diffuse scattering.  Our technique relies on two temporally separated optical pump pulses, where the second pulse suppresses or amplifies the coherences initiated by the first depending on the time delay between the pumps \cite{hase1996, decamp2001, misochko2013}.  We find that the patterns of suppression and amplification of the temporal oscillations in the diffuse scattering following the second pump as a function of oscillator frequency reveal that the phonons are squeezed in the cases of undoped, single-crystal Ge, GaAs, and InSb investigated here. This indicates that the short pump generates broad-band vibrational coherences in the form of pairs of correlated squeezed phonons spanning the BZ in all three materials studied. Our conclusion is based on the fact that the pump-pump delay required to suppress or amplify the temporal oscillations in $\braket{|u_{{\mathbf q}, \lambda}|^2}$ differs for a squeezed and coherent state of the same mode as illustrated in Fig. \ref{fig:f1} and discussed further below.  

We treat the phonons as independent harmonic oscillators that are perturbed by the pump pulses. Consider the simple case of a monatomic lattice in one dimension. We write phenomenologically the Hamiltonian for the single phonon branch interacting with optical radiation,
\begin{widetext}
\begin{multline}
H = \sum_q H_0(q) + \sum_q H_1(q,t) + \sum_{q,q'} H_2(q,q',t) = \\
\sum_q \frac{1}{2} \left( p_q p_{-q} + \omega_q^2 u_q u_{-q}\right)   + \sum_q A_1(q) f_1(t) u_q + 
\sum_{q,q^\prime} A_2(q,q^\prime) f_2(t) u_q u_{q^\prime}.
\label{eq:hamiltonian}
\end{multline}
\end{widetext}
Here $H_0$ is the Hamiltonian for the unperturbed phonons in the harmonic approximation, and $H_1$ and $H_2$ represent the interaction of an optical field with single and pairs of phonons, respectively. $u_q$ and $p_q$ represent the normal mode coordinates and their conjugate momenta, respectively. In the case of non-resonant Raman processes $A_1(q)$ and $A_2(q,q^\prime)$  
are proportional to the first and 
second-order Raman susceptibility, respectively, and $f_1(t) = f_2(t)$ is the pump pulse intensity.
\begin{figure*}
\centering
\includegraphics[width=\textwidth]{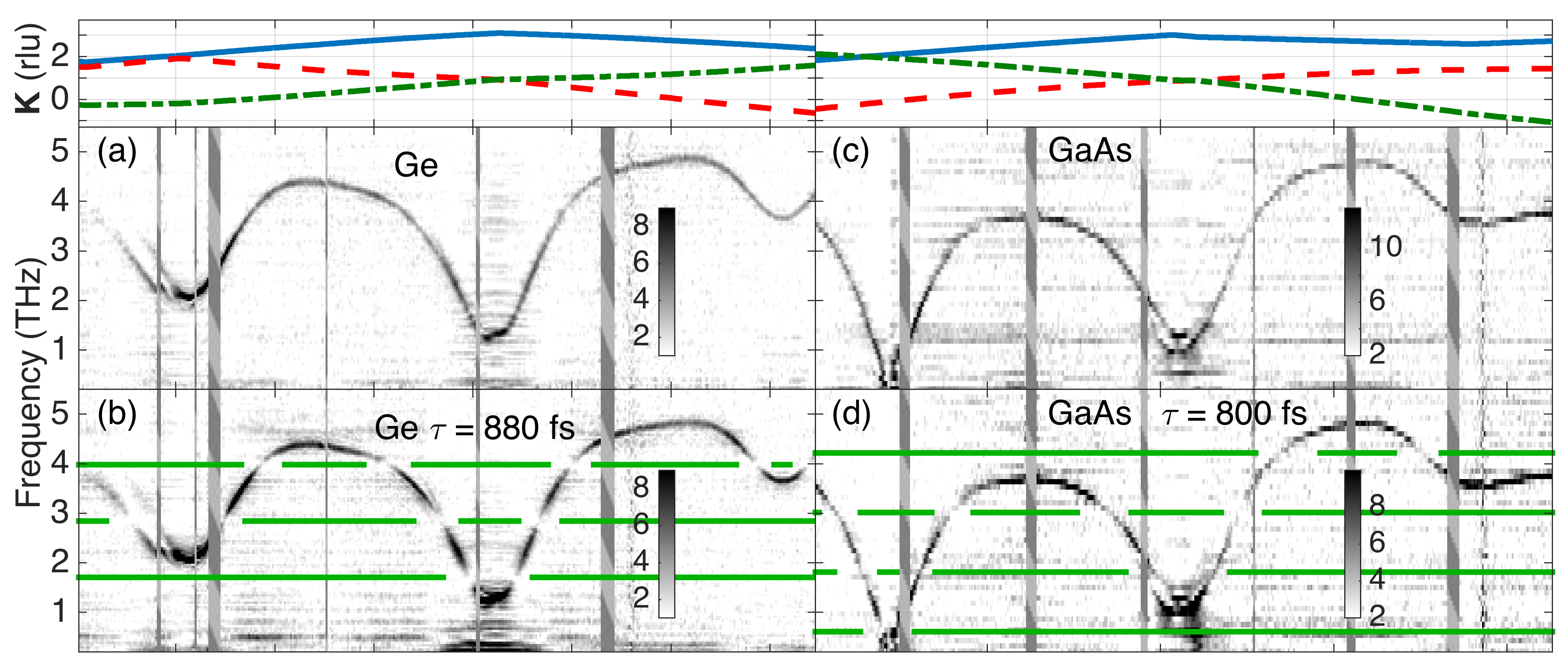}
\caption{(color online) {Dispersion of $\braket{|u_{\bf{q}, \lambda}|^2}$, matching twice the transverse-acoustic branches, obtained by Fourier transform of temporal oscillations in the diffuse x-ray intensity following single and double optical pump excitation.} (a) and (c) are with a single pump pulse while (b) and (d) are with two pumps {where} the pump-pump delay {is} labelled as $\tau$. The green horizontal lines in (b) and (d) show frequencies which should be stopped by the second pulse for squeezed phonons. The colorscale (in arbitrary units) in (d) has been {adjusted} to emphasize the contrast between frequencies suppressed and enhanced by the second pump. In the top plots, solid blue, dashed red, and dot-dash green curves label the three components of the scattering vector h, k, and l, respectively, in reciprocal lattice units . Vertical hashed bars are regions with no detector pixels.
}
\label{fig:f2}
\end{figure*}

By symmetry $A_1(q)$ is zero for high wavevector in an ordered crystal, leaving $H_2$ the leading order interaction. Approximating the wavevector of the  optical pump light as zero, momentum conservation imposes that $A_2$ is non-zero only when $q^\prime=-q$. The disordered case, in contrast, can have finite $A_1$ even at high $q$ \cite{carles1982}.
To see how to distinguish these processes, consider how the pump pulses affect a single high-wavevector mode for which $H_1$ is non-zero. Utilizing the arguments above, noting that $u_{-q}= u_q^*$ and $p_{-q} = p_q^*$ \cite{ziman1972}, and treating $u_q$ and $p_q$ as classical variables, we obtain the equation of motion for a single oscillator:
\begin{multline}
\ddot{u}_q = -\omega^2 u_q - A_1(q) f_1(t) - 2 A_2(q, -q) f_2(t) u_q.
\label{eq:eqmotion}
\end{multline}

For simplicity we assume the applied force is a delta function in time, and that the oscillators are initially in a thermal distribution. We can think of this as a distribution of classical oscillators, and in this context $\braket{|u|^2}$ denotes the ensemble average of $|u|^2$. Neglecting $H_2$, the equation of motion contains an impulsive force from $H_1$ which displaces the distribution along the momentum ($p$) axis to generate a coherent state as illustrated in Fig. \ref{fig:f1} (a). In contrast, the impulsive force derived from $H_2$ is proportional to the oscillator displacement, $u$. Thus if we neglect $H_1$ the right +$u$ side of the phase space distribution is pushed in the +$p$ direction, while the $-u$ side is pushed in the $-p$ direction, giving rise to a squeezed state as shown in Fig. \ref{fig:f1} (b). %Similar arguments hold for for (a) and (b) where the initial displacements are along $u$. 
{Although in this example we have considered only impulse-type excitation, displacive excitation is also possible, leading displacements along the $u$ axes \cite{zeiger1992}. Thus the phase of the oscillations in $\braket{|u|^2}$ could take any value depending on the degree of impulsive and displacive excitation.} 

Figure \ref{fig:f1} shows that $\braket{|u|^2}$ oscillates at \emph{twice} the phonon frequency for \emph{both} the squeezed and coherent cases. 
Clearly a measurement of the frequency of oscillations in $I({\bf K},t)\propto\braket{|u_{{\bf q}={\bf K}-{\bf G}}|^2} $  following single pump excitation is not sufficient to distinguish these two dynamic states of the oscillator. 

In Fig. \ref{fig:f1} we also illustrate how coherent control of the oscillator using two identical pump pulses could distinguish the true nature of the oscillator dynamical state in a second-order measurement of the phonon displacement. We start with a pair of equal-amplitude pulses separated by a fixed delay $\tau$. %From Fig. \ref{fig:f1}, 
We see that the coherent and squeezed motion are affected differently: to completely suppress the coherent phonon of natural frequency $\omega$ the second pulse has to arrive at delays $\tau = (2n+1)\pi/\omega$ for integer $n$ (i.e. odd multiples of half the oscillator period). On the other hand, to suppress the dynamics of the squeezed state of the same oscillator the second pump has to arrive at $\tau = (2n+1)\pi/2\omega$ (i.e. odd multiples of $\frac{1}{4}$ of the oscillator period). Also note that suppressing one case leads to enhancing the other because of the $\frac{1}{4}$ cycle phase difference. This method yields a clear way to distinguish pure coherent versus pure squeezed states based on which frequencies are suppressed or enhanced for a given $\tau$. For a general squeezed-coherent state, there is no $\tau$ which can completely suppress the oscillations.

Experiments were carried out at the X-ray Pump Probe (XPP) instrument \cite{xppref} of the Linac Coherent Light Source (LCLS) with a photon energy of 9.5 keV using a diamond double-crystal monochromater \cite{monoref}. The x-ray pulses were less than 50 fs in duration and contained  $\sim$10$^9$ photons at a repetition rate of 120 Hz. Optical pulses of $\sim$65 fs duration were provided by a multipass Ti:sapphire amplifier centered at a wavelength of 800 nm. Delay between optical and x-ray pulses was controlled electronically and measured using the XPP timing tool, leading to an overall time resolution $\sim$80 fs \cite{timingtool}. A Mach-Zehnder interferometer was used to control  the delay between two collinear optical pump pulses of equal pulse energy. A motorized shutter in one of the interferometer arms allowed control of single and double pump excitation. Scattering measurements were performed in a reflection geometry with x-ray grazing angles varying from 0.45--1.0 degree to match the optical and x-ray penetration depths for the various samples. The optical laser was 0.5 degrees less grazing than the x rays and p-polarized. The x-ray and optical laser beam cross sections were 200 $\times$ 15 and 600 $\times$ 90 $\mu$m$^2$, respectively. The samples were commercial single-crystal wafers of undoped Ge, GaAs, and InSb. Samples were mounted in a He-purged environment to minimize air scattering. Scattered x rays were collected using the Cornell-SLAC pixel array detector (CSPAD) on a shot-by-shot basis \cite{cspad}. In each case, the sample orientation and detector position is fixed, and thus we probe  a two-dimensional slice of reciprocal space.  Each pixel on the CSPAD detector corresponds to a small spread in momentum transfer and thus crystal momentum set by the scattering geometry and the sample. 

Figure \ref{fig:f2} shows a portion of the measured dispersion with single and double-pump excitation in Ge and GaAs obtained using analysis described previously \cite{zhu2015, trigo2013}.  Here we have   selected a one-dimensional path spanning multiple zones through the two-dimensional projection in reciprocal space.  The dispersion is obtained by plotting the temporal Fourier transform of  the differential change in $I(\mathbf{K},t)$ following the excitation by the pump pulse(s) for each pixel. The observed dispersions match twice the frequency of the transverse-acoustic phonons in all three materials studied, as expected for a measurement of $\braket{|u_{\mathbf{q}, \lambda}|^2}$ and consistent with refs \cite{trigo2013,zhu2015}.  As seen in Fig. \ref{fig:f2}, the optical laser produces phonon coherences which span the BZ and have a continuum of frequencies. Thus a fixed pump-pump delay is in principle all that is necessary to distinguish squeezed from coherent motion assuming a single generation mechanism dominates. The second pulse excitation will maximally suppress the amplitude of oscillation only of those modes for which $\omega \tau \sim (2n+1)\pi/2$ if they are squeezed modes, or $\omega \tau \sim (2n+1)\pi$ if they are coherent modes. 
The horizontal lines show frequencies which should be stopped by the second pulse if the phonons are squeezed. One clearly sees a decrease in the Fourier intensity at these frequencies when compared to the single-pump. This is consistent with the high-wavevector phonons being squeezed and thus  generated by scattering of phonon pairs as represented by the interaction term $H_2$ in Eq. (\ref{eq:hamiltonian}).

\begin{figure}
\centering
\includegraphics[width=3.375in]{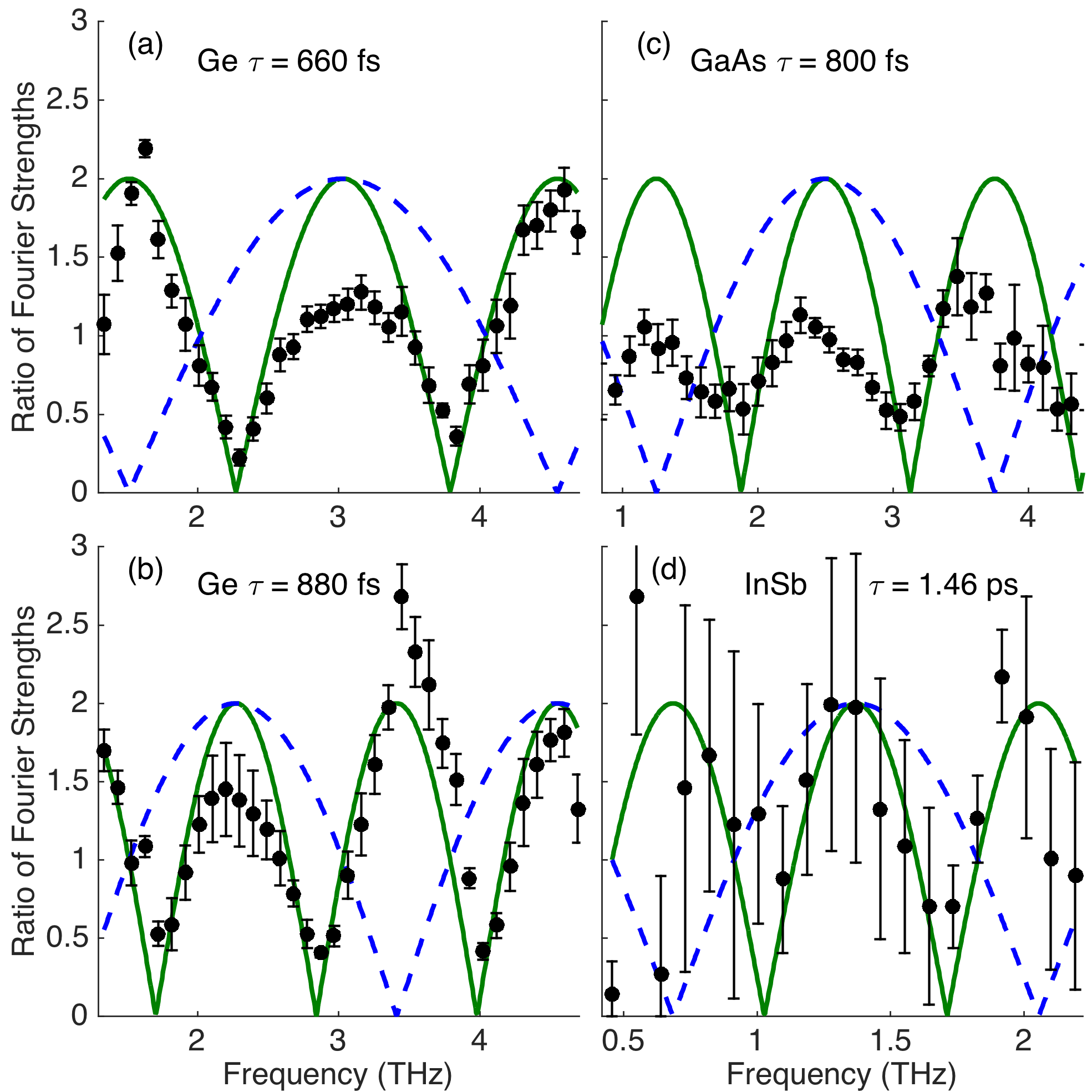}
\caption{(color online) Oscillation amplitude of phonon-phonon correlation following two-pulse excitation. The data is binned according to frequency and normalized to single-pulse excitation amplitude. The pump-pump delay is labelled as $\tau$. Solid green and dashed blue curves show expected trends for impulsive excitation for squeezed and coherent states, respectively with no free parameters. 
}
\label{fig:f3}
\end{figure}

To more quantitatively test the generation mechanism, we compared the temporal Fourier transform amplitude of the scattered signal after single and double-pump excitation as a function of frequency.% (Fig. \ref{fig:f3}).    
Within the two-dimensional slice of reciprocal space probed, a mask was applied to exclude detector pixels with low scattering signal. Each unmasked pixels was assigned a frequency  according to the maximum of their Fourier transform for the single pump excitation, and data from pixels with the same frequency were combined. 
In Fig. \ref{fig:f3}, we plot the integrated Fourier peak for the double-pump excitation normalized to the single pump for each frequency bin. 
The error bars represent the standard mean error obtained when applying the same analysis to independent repeated measurements, while the points represent the average. Also shown is the calculated ratio for squeezed (green solid line) and coherent (blue dashed line) phonons assuming a delta function excitation. In all cases the data matches the trends expected for squeezed phonons much more closely than those expected for coherent phonons. We conclude that the optical pump primarily generates pairs of phonons with nearly equal and opposite momenta.

We emphasize that here we observe oscillations only at twice the phonon frequency. 
X-ray diffuse scattering can also detect $\braket{\mathbf{u_{q}}}$ in first-order in the presence of significant elastic scattering at $\bf{K}=\bf{G}+\bf{q}$. In this case the time domain signal corresponds to a heterodyne detection between the elastic and inelastically scattered x rays at a given momentum transfer and will thus oscillate at the phonon frequency for a coherent state \cite{reis2007}.  In the case of disorder, we expect that the first-order continuum will be generated preferentially at the same momentum transfers that the elastic diffuse scattering is large.  Therefore we expect a signal at both the fundamental and second harmonic with their relative magnitudes depending on the details of the excitation and detection. 

Our method for distinguishing squeezed from coherent phonons is valid for any time-resolved measurement of the second-order correlations in the atomic-displacements. In the case of ultrafast optically-excited Ge, GaAs and InSb, we confirm that the observed oscillations at twice the phonon frequency in the diffuse scattering signal originate from squeezed phonon modes spanning the Brillouin-zone.  This is consistent with the coherences in the second-order equal-time correlations, $\braket{|u_{\mathbf{q}, \lambda}|^2}$, being generated by a second-order process that creates correlated pairs of phonons with near equal and opposite momentum. This type of generation process is generally allowed by symmetry selection rules, making this method a broadly applicable technique to study collective excitations in solids. 

{This work is primarily supported by the U.S. Department of Energy, Office of Science, Office of Basic Energy Sciences through the Division of Materials Sciences and Engineering under Contract No. DE-AC02-76SF00515.} Measurements were carried out at the Linac Coherent Light Source, a national user facility operated by Stanford University on behalf of the U.S. Department of Energy, Office of Basic Energy Sciences. Preliminary measurements were performed at the BioCARS at the Advanced Photon Source.  The Advanced Photon Source is  supported by the U.S. Department of Energy, Basic Energy Sciences, Office of Science, under Contract No. DE-AC02-06CH11357. Use of BioCARS was also supported by the National Institutes of Health, National Institute of General Medical Sciences grant 1R24GM111072. JNC gratefully acknowledges financial support from the Volkswagen Foundation. SF acknowledges support by Science Foundation Ireland under grant 12/IA/1601. DAR acknowledges the SFI Walton Visitor award, Science Foundation Ireland grant 11/1W/I2084 and discussions with Roberto Merlin.

\bibliography{ref.bib}{}

\end{document}